\documentclass[preprintnumbers,aps,twocolumn,floatfix]{revtex4} 
 
\usepackage{bm} 
\usepackage[english]{babel} 
\usepackage[latin1]{inputenc} 
\usepackage[dvips]{graphicx} 
\usepackage{amsmath} 
 
\renewcommand{\d}{{\rm d}} 
\newcommand{\e}{{\rm e}} 
\newcommand{\imai}{{\rm i}} 
 
\usepackage{epsfig} 
 
\begin{document} 
\preprint{Accepted by Applied Physics Letters
  and scheduled for Jan. 2005, see \url{http://apl.aip.org/apl/}}

\keywords{Quantum Cascade Laser, Linewidth, intersubband transition} 
\pacs{73.63.-b, 42.55.Px,78.67.-n} 


\title{Self-Consistent Theory of the Gain Linewidth for 
Quantum-Cascade Lasers} 
 
\author{F.~Banit} 
\author{S.-C.~Lee}  
\author{A. Knorr} 
\affiliation{Institut f{\"u}r Theoretische Physik, 
Technische Universit{\"a}t Berlin, D-10623 Berlin, Germany} 
\author{A.~Wacker} 
\email{Andreas.Wacker@fysik.lu.se} 
\affiliation{Fysiska Institutionen, Lunds Universitet, Box 118, 22100 Lund,  
Sweden} 
 
\begin{abstract} 
The linewidth in intersubband transitions can be significantly 
reduced below the sum of the lifetime broadening for the involved states, 
if the scattering environment is similar for both states.  
This is studied within a nonequilibrium Green function approach here. 
We find that the effect is of particular relevance for a recent, 
relatively low doped, THz quantum-cascade laser.
\end{abstract} 
 
\maketitle 
 
Material gain engineering in semiconductor quantum wells is a major topic  in
the physics of semiconductor laser emission \cite{Hader1}. In particular, the
position of the peak gain and the linewidth of the  active medium are of
importance for applications in telecommunication or short pulse
generation. Whereas the gain properties of semiconductor interband quantum
well lasers  are well understood \cite{Chow1}, intersubband emitters, such
as quantum-cascade lasers (QCLs) \cite{FAI94a}, are still a subject of
intensive research. Major differences of QCLs in comparison to interband
lasers  result from the similar in-plane band curvature  for the electronic
transitions and the low energy collective excitations involved in the photon
emission. Besides model calculations \cite{LEE02a,PER04} for  a standard QCL
design \cite{SIR98}, a detailed many-particle study of the gain linewidth of
intersubband lasers for the broad variety of QCL structures available
\cite{Gmach1,KOE02,KUM04} is still missing. The purpose of this Letter is to
evaluate a self-consistent theory for the influence of the electronic
scattering mechanism  on the gain linewidth of QCLs and to clarify its impact
on the description of QCL structures ranging from the infrared to THz-regime.
 
For interband lasers the many-particle interactions are well described on
the level of the second order Born approximation \cite{Chow1}. Due to the
expansion of the scattering matrix  in wavenumber states, the linewidth of the
transitions is determined by diagonal (proportional to the inverse lifetime)
and non-diagonal dephasing mechanisms \cite{Lind1,Rossi1}. The consistent
description of the detailed interplay of  diagonal and non-diagonal scattering
mechanisms \cite{Chow2}  results in a proper prediction of the gain shape.
In particular spurious absorption below the band  gap could be eliminated from
the theory \cite{Hugh1}. Our aim is to evaluate similar information for the
gain in the intersubband gain of QCL structures. So far, only simple two band
intersubband emitters \cite{Ines1,Li1}  have been investigated on this
self-consistent level.
 
Here, we bridge the gap to real QCL devices, and explore the influence of a
self-consistent description of the electron-phonon and electron-impurity
scattering on the gain linewidth. Although excluded in the  present study,
electron-electron scattering can be treated on the same footing \cite{Ines1}.
Our results are illustrated for two different samples: (A) the terahertz QCL
reported in Ref.~\cite{KUM04}, (B) a midinfrared  QCL reported in
Ref.~\cite{SIR98}. We demonstrate, that the different behavior can be
understood by the impact of non-diagonal dephasing on the linewidth. Finally,
we give general conditions, when non-diagonal dephasing is important and
thus a full quantum theory of the gain is indispensable.
 
\begin{figure}
    \includegraphics[width=6.2cm]{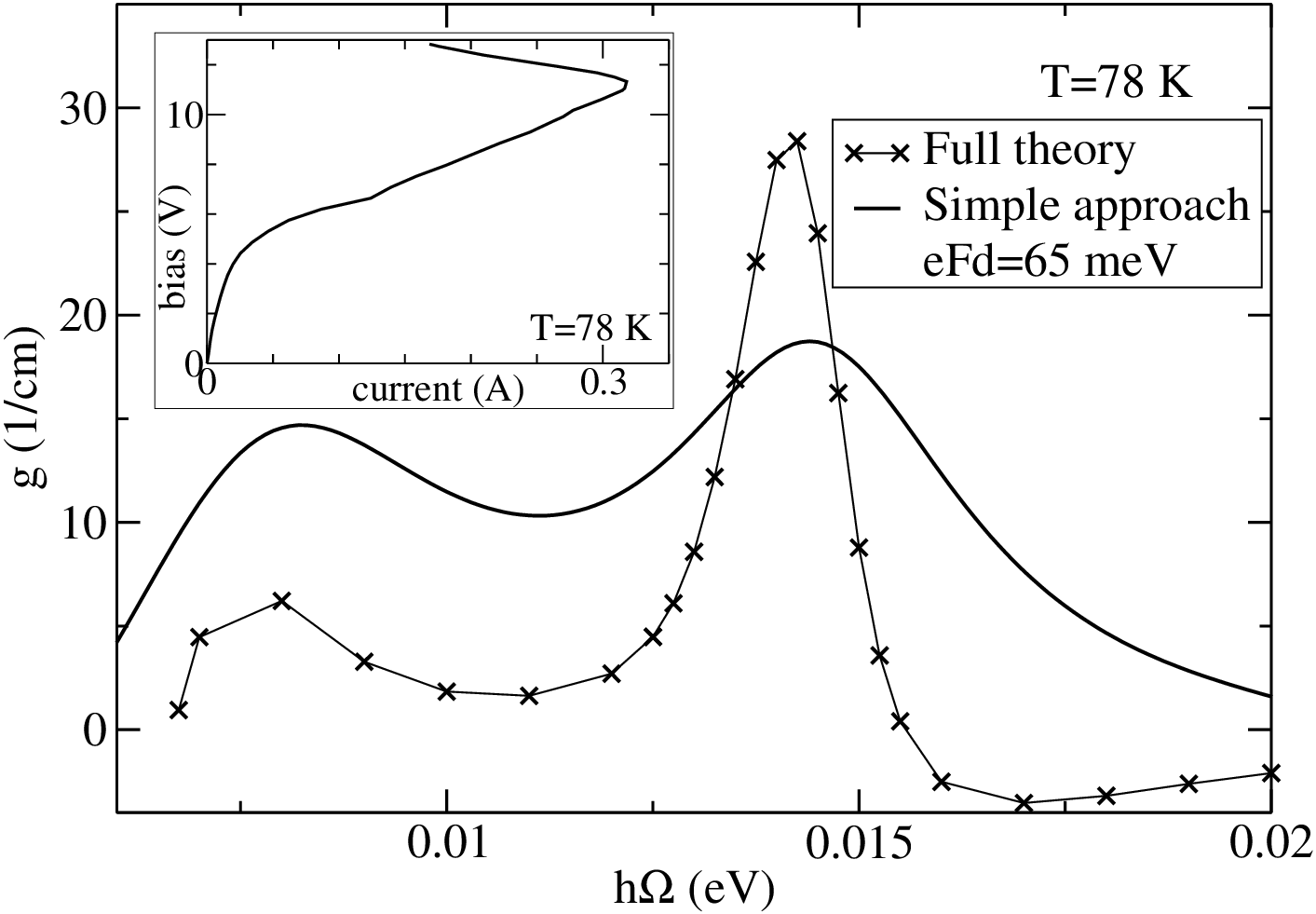} 
    \caption{Calculated gain at an operating bias 
of 11.5 V for sample A  (Ref.~\cite{KUM04}). The inset shows the  
current-voltage characteristic where the bias equals 177 times the  
bias drop per period $Fd$ and the current is evaluated for an  
area of 54000$\mu$m$^2$.}   
\label{FigGainKum} 
\end{figure}

First, we evaluate the nonequilibrium electron occupations with
nonequilibrium  Green functions \cite{LEE02a} using nominal sample
parameters. Here, however, we extend the model in two ways: (i)
Scattering at ionized impurities is treated microscopically 
using Debye screening by a fictitious 
homogeneous 3D electron gas matching the average doping density.
(ii) We include the
non-diagonal elements in the self-energies, so that Eqs.~(6-8) of
Ref.~\cite{LEE02a} take the form:
\begin{multline}
\Sigma_{\alpha_1\alpha_2}^{{\rm ret},{\rm imp}}({\bf k},E) = \\
\sum_{\beta_1\beta_2,{\bf k'}} 
\underbrace{\langle V_{\alpha_1 \beta_1}^{\rm imp}( {\bf k - k'}) 
V_{\beta_2 \alpha_2}^{\rm imp}( {\bf k' - k})\rangle}_{=\hbar^2 
\gamma_{\alpha_1\alpha_2\beta_1 \beta_2}/(Am_c) } 
\,G_{\beta_1 \beta_2}^{\rm ret}({\bf k'},E)~, 
\label{EqSigRough} 
\end{multline} 
and similarly for the electron-phonon scattering processes ($A$ denotes the
normalization area and $m_c$ is the effective mass). The values of
$\gamma_{\alpha_1\alpha_2\beta_1 \beta_2}$ are taken for a typical fixed
momenta $|{\bf k}|,|{\bf k'}|$, while the integration over $\angle({\bf
k},{\bf k}')$  is performed. 
The current voltage characteristic is evaluated along the lines  of
Ref.~\cite{LEE02a} and is shown in 
the inset of Fig.~\ref{FigGainKum} for the THz-QCL A\cite{Sampl1}. It
agrees excellently with the data presented in Fig. 2 of Ref.~\cite{KUM04}.
 
\begin{figure}
   \includegraphics[width=6.2cm]{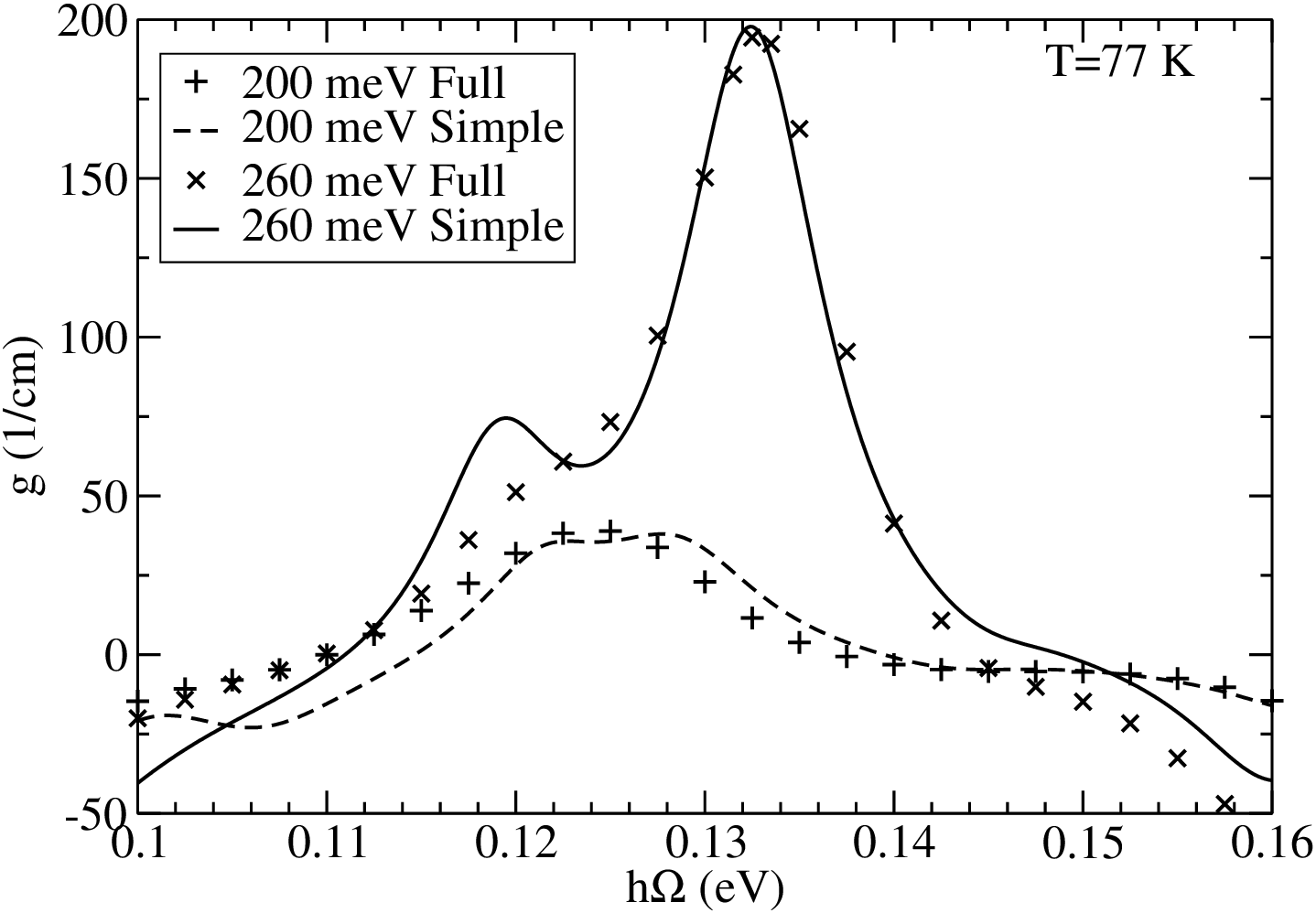} 
\caption{Calculated gain for sample B (Ref.~\cite{SIR98}).\hfill} 
\label{FigGainSir}
\end{figure}

Next, we evaluate the gain according to Ref.~\cite{WAC02a}. The material gain
coefficient $g(\Omega)$ reads as a function of the frequency $\Omega$
\cite{Chow1}:
\begin{equation} 
g(\Omega)=-\Re\left\{\frac{ \delta J} {\delta
    F(\Omega)c\epsilon_0\sqrt{\epsilon_r}}\right\}~.
\label{Eqgain} 
\end{equation} 
where $\epsilon_r=13$ is the dielectric constant. The change in the current
density $\delta J$ for a perturbative electric field  $\delta F(t)=\delta
F(\Omega)\e^{-\imai\Omega t}$ is given by \cite{WAC02a}:
\begin{equation} 
\delta J = \frac{e}{d\hbar} 
\int\frac{\d E}{2\pi}\frac{2}{A}\sum_{\bf k}\sum_{i,j} 
z_{ji}(E_j-E_i)\delta G^<_{ij}({\bf k},E) 
\label{EqDeltaJ} 
\end{equation} 
where  the states $\Psi_i(z)$ are the  eigenstates of the heterostructure
potential including  the self-consistent Hartree field, $z_{ij}$ is the dipole
matrix element, and $d$ is the period of the structure. The field-induced
changes $\delta G^{<}_{ij}$ of the Green functions are evaluated
self-consistently including the  non-diagonal elements of the self-energies in
the $\delta \Sigma$ terms ({\em full theory}). The results are given in
Figs.~\ref{FigGainKum},\ref{FigGainSir} (crosses) for the two samples
considered here. For comparison, we have evaluated the gain by the {\em simple
approach} (using lifetime broadening only):
\begin{equation}\begin{split}
g_{\rm simple}(\Omega)=&\sum_{i,j}
\frac{e^2|z_{ij}|^2(E_i-E_j)(n_i-n_j)}{2d\hbar c\epsilon_0\sqrt{\epsilon_r}}\\
&\times 
\frac{\Gamma_i+\Gamma_j}{(E_i-E_j-\hbar \Omega)^2+(\Gamma_i+\Gamma_j)^2/4}
\label{EqLorentz}
\end{split}\end{equation}
Here, we evaluate the level energies $E_i$ and their respective 
widths $\Gamma_i$ (FWHM)  
from the spectral functions $-2\Im\{G^{\rm ret}_{ii}(k=0,E)\}$ 
and the densities are  
given by $n_{i}=\sum_k\int\d E \Im\{G^{<}_{ii}(k,E)\}/(\pi A)$. 
Eq.~(\ref{EqLorentz}) can be derived restricting to diagonal dephasing, 
see, e.g., Ref.~\cite{GEL96}, where effects due to different 
effective masses were studied, which are neglected here.
It has also been used in Ref.~\cite{LEE02a} except 
for the replacement $E_i-E_j=\hbar \Omega$ in the first line 
and the neglect of counter-rotating terms with $E_i<E_j$. 
Figs.~\ref{FigGainKum},\ref{FigGainSir} show, that the simple 
approach is in good agreement with the full theory 
for sample B, but exhibits a far too large width of the 
gain peak for sample A. This shows, that the simple (Lifetime)  
approach does not correctly describe the gain spectrum for all QCLs.

In the following we want to shed light into the physical 
mechanisms which cause the failure of the simple approach (\ref{EqLorentz}). 
For this purpose we assume that the Green functions $\tilde{G}_{ij}$ 
of the stationary state are diagonal in our basis of eigenstates
and given by their  generic forms 
\begin{eqnarray}
\tilde{G}_i^{\text{ret/adv}}({\bf k},E)&=&
\frac{1}{E-E_i-E_k\pm\imai \Gamma_i/2}
\label{EqGretsimple}\\
\tilde{G}_i^{<}({\bf k},E)&=&
\imai f_i(E)\frac{\Gamma_i}{(E-E_i-E_k)^2+\Gamma_i^2/4}
\label{EqGlesssimple}
\end{eqnarray}
Neglecting the $\delta\Sigma$-terms provides us with, see Ref.~\cite{WAC02a}: 
\begin{multline} 
\delta G^{<0}_{ij}({\bf k},E)= 
\tilde{G}_i^{\text{ret}}({\bf k},E+\hbar \Omega) 
(-e\delta F(\Omega)z_{ij})\tilde{G}_j^{<}({\bf k},E)\\ 
+\tilde{G}_i^{<}({\bf k},E+\hbar \Omega) 
(-e\delta F(\Omega)z_{ij})\tilde{G}_j^{\text{adv}}({\bf k},E)
\label{EqDeltaG0} 
\end{multline}
We define $\delta g_{ij}^{<0}(E)=\int_0^{\infty}
\d E_{k} \delta G^{<0}_{ij}({\bf k},E)$.
Extending the $E_k$ integration to $-\infty$ 
(which is possible only for $E>{\rm Max}\{E_i-\hbar\Omega,E_j\}$)
we find
\begin{equation}
\delta g_{ij}^{<0}(E)\approx 2\pi\imai
\frac{-e\delta F(\Omega)z_{ij}[f_j(E)-f_i(E)]}
{E_j+\hbar\Omega-E_i+\imai(\Gamma_i+\Gamma_j)/2}
\label{Eqg0}
\end{equation}
which justifies Eq.~(\ref{EqLorentz}) by inserting into 
Eqs.~(\ref{Eqgain},\ref{EqDeltaJ}).
Thus the simple approach can be related
to the neglect of $\delta\Sigma$-terms in our full theory.
 
\begin{figure} 
\includegraphics[width=8.5cm]{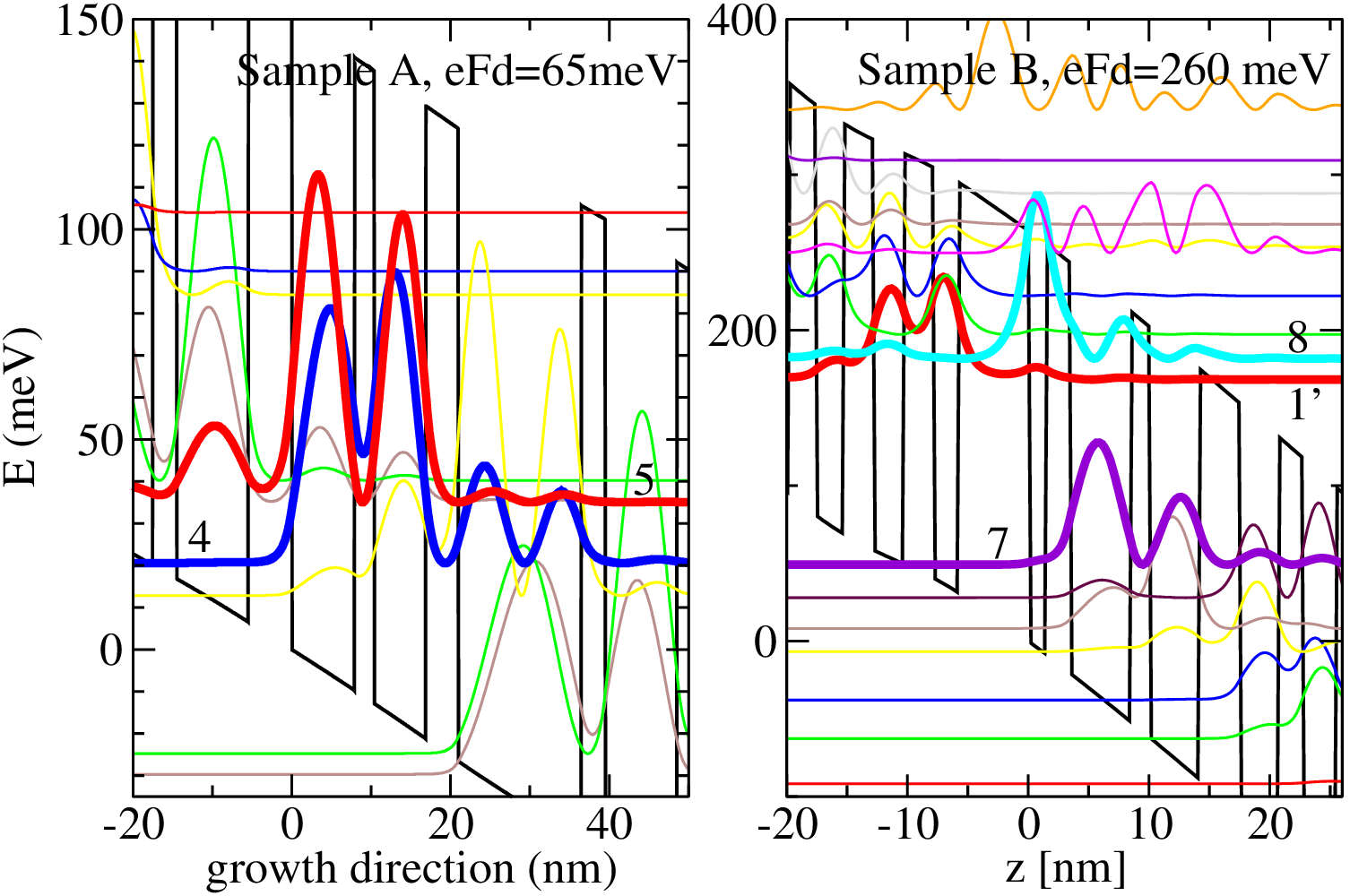} 
\caption{Band diagrams for samples A and B together with main  levels.\hfill} 
\label{FigWSfunc} 
\end{figure}

Now we want to examine the $\delta\Sigma^<$-terms (neglecting
$\delta\Sigma^{\rm ret/adv}$ for the purpose of simplicity in the analytic 
approach), which give:
\begin{multline}
\delta G^{<}_{ij}({\bf k},E)\approx\delta G^{<0}_{ij}({\bf k},E)\\
+\tilde{\bf G}_i^{\text{ret}}({\bf k},E+\hbar \Omega)
\delta \Sigma_{ij}^<(E)\tilde{\bf G}_j^{\text{adv}}({\bf k},E)
\label{Eqself1}
\end{multline}
For $i\neq j$, the dominant contribution \cite{InfoDominant} to
$\delta \Sigma_{ij}^{<,{\rm imp}}$ takes the form
\begin{equation}
\delta \Sigma_{ij}^{<,{\rm imp}}(E) = \gamma^{\rm imp}_{ijij}/2\pi
\int_0^{\infty}\d E_{k'} \delta G_{ij}^{<}({\bf k'},E)\, ,
\label{EqDeltaSigma}
\end{equation}
which gives the non-diagonal dephasing as a ${\bf k'}$ integral similar 
to density matrix theory \cite{Lind1,Rossi1,Chow2,Hugh1}.
Now we define
$\delta g_{ij}^<(E)= \int\d E_{k} \delta G_{ij}^<({\bf k},E)$.
Again extending the $E_k$ integration to $-\infty$, 
Eqs.~(\ref{EqGretsimple}-\ref{EqDeltaSigma}) give
\begin{equation}
\delta g_{ij}^<(E)\approx 2\pi\imai
\frac{-e\delta F(\Omega)z_{ij}[f_j(E)-f_i(E)]}
{E_j+\hbar\Omega-E_i+\imai(\Gamma_i+\Gamma_j-2\gamma^{\rm imp}_{ijij})/2}
\end{equation}
which has the form discussed in Ref.~\cite{AndoWidth}. We immediately see,
that the width of the gain peak is reduced. This relates to the line
narrowing due  to non-diagonal elements in the self-energies (or non-diagonal
dephasing, see also \cite{Ines1,Li1}). Impurity scattering contributes with
approximately  $\gamma^{\rm imp}_{iiii}$ to the full broadening $\Gamma_{i}$
of the states $i$. Therefore it is crucial to compare $\gamma^{\rm
imp}_{ijij}$ with  $\gamma^{\rm imp}_{iiii}$ and $\gamma^{\rm imp}_{jjjj}$.
From Eq.~(\ref{EqSigRough}) we find that $\gamma^{\rm imp}_{ijij}$ is of the
order of $\sqrt{\gamma^{\rm imp}_{iiii}\gamma^{\rm imp}_{jjjj}}$ if the
scattering potentials $V_{ii}^{\rm imp}$ and $V_{jj}^{\rm imp}$ are
correlated, i.e. both states $\Psi_i(z)$ and $\Psi_j(z)$ are exposed to an
identical scattering environment. In this case the transitions are less
affected by the scattering, resulting in a narrowing of the gain
feature. Indeed we find for the dominating 4-5 transition  (see
Fig.~\ref{FigWSfunc}) for sample A $\gamma^{\rm imp}_{4545}= 2.2$ meV,
$\gamma^{\rm imp}_{4444}= 2.7$ meV, and  $\gamma^{\rm imp}_{5555}= 2.2$
meV. This means both scattering environments are highly correlated, resulting
in the strong narrowing in the gain visible in Fig.~\ref{FigGainKum}.  In
contrast the main gain peak at $eFd=260$meV in sample B can be attributed
to the 7-8  transition with  $\gamma^{\rm imp}_{7878}=1.1$ meV,
$\gamma^{\rm imp}_{7777}=1.9$ meV, and  $\gamma^{\rm imp}_{8888}=8.3$
meV. Thus, correlations in the scattering  environment are weak and the simple
approach works well, see Fig.~\ref{FigGainSir}.

The difference between the two designs can be understood by considering  the
nature of the scattering potentials. Sample A is low-doped, corresponding to
a screening length of 30 nm, and the differences between  the  wavefunctions 4
and 5 occur on a smaller scale, see Fig.~\ref{FigGainSir}. In contrast the
screening length is 8 nm for the high-doped sample B. Thus the scattering
potential is much more local for sample B causing significant differences in
the scattering environment for both levels. This effect is strengthened by
the fact that interface roughness (with $\delta$-function potentials) is
also of relevance for sample B.
 
Note that the analytical motivation given above reproduces the right trend,
but cannot be used for quantitative analysis as: (i) Phonon scattering shows
also some line narrowing due to correlations in the potentials, which is
contained in the full theory. (ii) The
influence of $\delta \Sigma_{ij}^{\rm ret/adv}(E)$ has been neglected. (iii)
Significant non-diagonal elements in $G_{ij}^<(E)$ are present even in the
basis of energy eigenstates.
 
In {\em conclusion}  we have shown that the simple approach (\ref{EqLorentz})
for QCL gain is not reliable if the states involved in the lasing transition
are exposed to the same scattering environment. Non-diagonal dephasing then
becomes important and narrows the linewidth below the sum of the individual
widths of the levels (lifetime broadening). This requires a full quantum
kinetic description of the gain in QCLs.

This work was supported by Deutsche Forschungsgemeinschaft (DFG) and the
Swedish Research Council.

\end{document}